\documentclass[conference]{IEEEtran}
\IEEEoverridecommandlockouts
\usepackage{cite}
\usepackage{amsmath,amssymb,amsfonts}
\usepackage{algorithmic}
\usepackage{graphicx}
\usepackage{textcomp}
\usepackage{xcolor}
\def\BibTeX{{\rm B\kern-.05em{\sc i\kern-.025em b}\kern-.08em
    T\kern-.1667em\lower.7ex\hbox{E}\kern-.125emX}}

\begin{document}

\title{Job Class Thermal Intent Aware Liquid Cooling Allocation for AI Data Centers}

\author{
    \IEEEauthorblockN{Krishna Chaitanya Sunkara}
    \IEEEauthorblockA{\textit{AI Data Center Engineering}\\
    \textit{Oracle}\\
    Raleigh, USA \\
    ORCID:0009-0009-6159-4280}
}

\maketitle

\begin{abstract}
GPU-dense AI data centers need to run on liquid cooling as air simply cannot shed the heat at these power densities. Yet the cooling loops themselves are blind to what workloads are about to run; they crank up flow only after a sensor catches a temperature climb, which can take 30 to 50 seconds. We built Job-Class Thermal Intent (JCTI) to close that window. The scheduler already knows a job is coming and what class it belongs to; JCTI feeds that information straight to the cooling controller so it can stage coolant before the heat shows up. We pulled the thermal signatures for each job class out of MLPerf GPU power traces and tuned arrival patterns against Alibaba cluster data. Over 120 paired Monte Carlo trials the numbers come out to 56.4\% fewer thermal violations and 60.2\% less cumulative overshoot than a straight PI loop. As AI data centers evolving towards gigawatt grid loads with highly fluctuating power swings, thermally-aware scheduling reduces sudden demand and improves load prediction in grid side. Cooling and scheduling have been running as two separate systems for years despite each one knowing something the other needs, JCTI wires them together.
\end{abstract}


\section{Introduction}
GPU-dense AI deployments have pushed rack-level power densities well beyond what forced-air systems can handle. NVIDIA H100 GPUs alone dissipate up to 700W each, and at eight per server the numbers add up quickly past 60kW per rack. ASHRAE TC 9.9 guidelines make clear that at these densities, air cooling is no longer a viable path forward \cite{ashrae2021}. Liquid cooling has consequently become the only real option for any serious AI cluster build-out. What has changed less rapidly is how that liquid is actually controlled once installed. Training runs sustain high, relatively steady power draws punctuated by checkpoint flushes, while inference workloads swing hard with every batch-size adjustment or request surge \cite{dean2013tail}. Analytics jobs sit somewhere in between, their power profiles shaped by whatever data they happen to be processing at the moment.

Today's cooling controllers sit and wait for a temperature reading to creep up before doing anything about it. A sensor catches the rise, the controller works out a new valve position, and the valve moves, all of which eats up somewhere around 30 to 50 seconds in a real installation \cite{shahi2022flow}. That latency gap is fine when workloads change slowly, but inference jobs can go from idle to full power in two or three seconds \cite{kandala2020power}. By the time the feedback loop catches up, the damage is already done and GPU throttling has kicked in. AI data centers emerging as gigawatt grid loads. This demand can cause power swings in hunreds of kilowatts. Thermally-aware workload scheduling reduces demand fluctuations and improves grid-side prediction.
Lu et al. recently looked at placing jobs with awareness of cooling headroom \cite{lu2025thermal}, essentially treating the cooling system as a constraint the scheduler has to work around. Zhao et al. took the other side and built a predictive controller that adjusts supply based on measured thermal state across zones \cite{zhao2025mpc}, but without any knowledge of what workloads are actually going to arrive. Neither line of work ever passes information about incoming jobs down to the valve controllers, which is where the real opportunity sits.

Our goal here is to let the cooling system know what is coming before it arrives. Rather than reacting to temperature, we use the scheduler's own metadata, job type, GPU count, expected duration, and crucially the short window of time between when a job is dispatched and when it actually starts running, to pre-position coolant. We call this Job-Class Thermal Intent, or JCTI. Concretely, we derive stochastic power-profile models per job class from MLPerf GPU measurements \cite{mlperf2023} and calibrate arrival statistics to Alibaba production traces \cite{gong2010press}. On top of that we define a readiness metric that captures both how fast a rack can stabilize after a new job lands and the probability of a throttling-inducing violation over a forward horizon. The thermal plant model accounts for multiple racks sharing a single CDU with hard flow limits, and the controller combines feedforward pre-positioning with light feedback correction. Everything is validated over 120 paired Monte Carlo trials with paired t-tests and ablations on dispatch notice availability and job-class accuracy.

\section{Thermal Intent Architecture and Derivation}
\begin{figure}[t]
    \centering
    \includegraphics[width=\columnwidth]{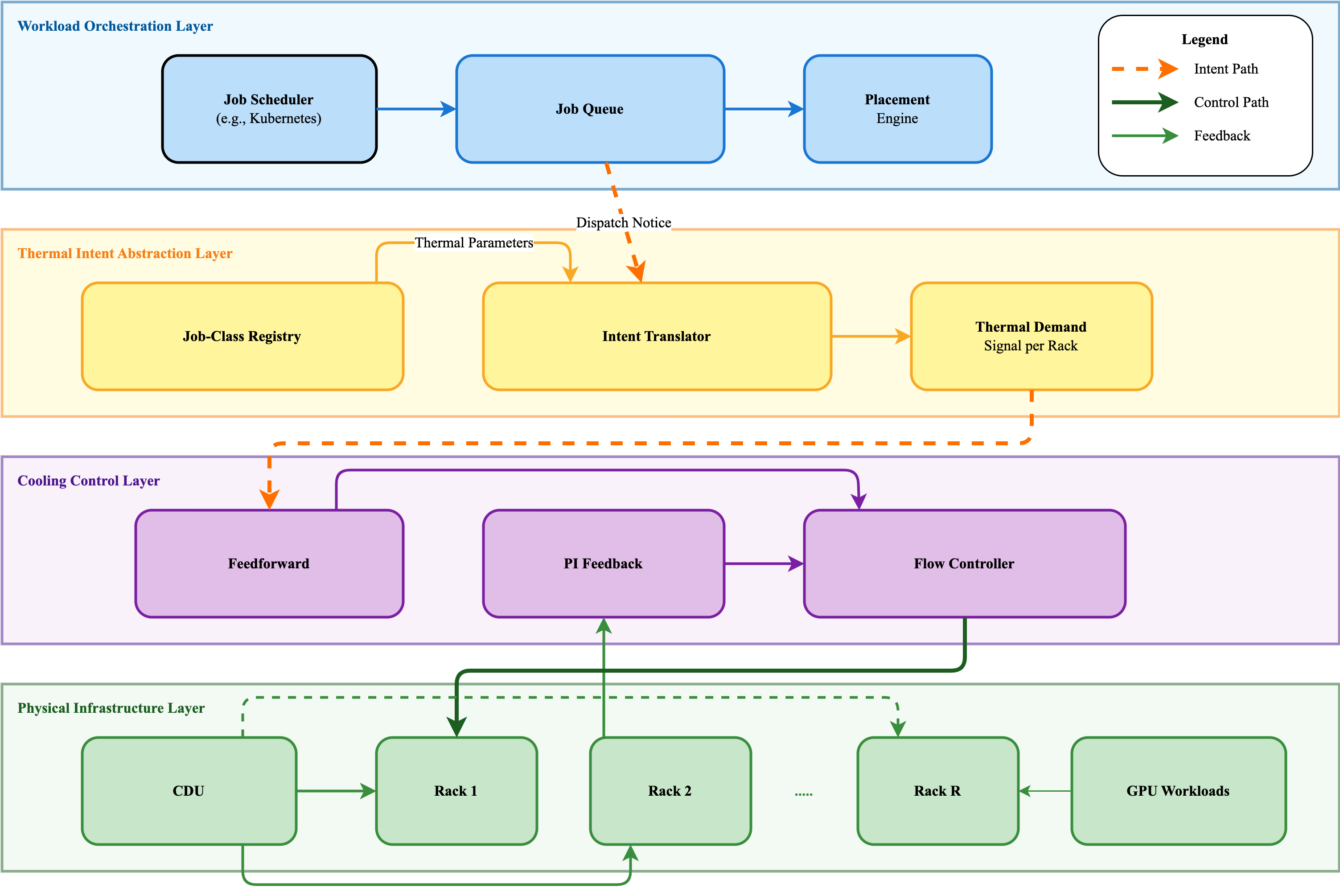}
    \caption{JCTI architecture showing four-layer design: workload orchestration extracts job-class and dispatch timing, thermal intent layer translates to rack-specific demand signals $D_r(t)$, cooling control combines feedforward and PI feedback paths, and physical infrastructure actuates coolant flow $m_r(t)$ across R racks sharing a CDU.}
    \label{fig:architecture}
\end{figure}

The system is built from four pieces that work together to get scheduler knowledge down to the valve level. The scheduler or orchestrator spits out job descriptors
\begin{equation}
j \equiv \{c_j,\; \pi_j,\; \hat{T}_j,\; \Delta_j,\; \rho_j\}
\end{equation}
where $c_j$ is job class (training, inference, analytics), $\pi_j$ says how many GPUs and what type, $\hat{T}_j$ is the expected run duration, $\Delta_j$ is how far ahead the dispatch notice arrives before execution, and $\rho_j$ sets priority for when resources get tight. None of that requires changing how Kubernetes or Slurm actually works, it is all metadata that is already sitting there.

The system architecture Fig.~\ref{fig:architecture} bridges workload orchestration and cooling actuation.
A thermal intent service keeps a library of JCTI parameter sets $\Theta_{c_j}$, one per job class, and uses them to spit out short-horizon power and heat forecasts whenever a new job descriptor comes in. Keeping that lookup separate from the scheduler means we can update the thermal models from live telemetry without touching the orchestrator. The two subsystems stay independent and can be tuned on their own schedules.

The cooling allocation controller takes those forecasts plus the current rack temperatures and computes per-rack flow targets $\dot{m}_r(t)$, respecting valve rate limits, capacity bounds, and the shared CDU constraint. It runs feedforward pre-positioning driven by dispatch notice and layered with feedback correction on the actual temperature reading. A safety layer sitting underneath enforces hard limits consistent with the flow control hardware characterization in \cite{shahi2022flow}, no unsafe slew rates, no flow swings that could hammer the hydraulics. We pulled the thermal signatures from two real data sources. MLPerf benchmark submissions give us GPU-level power measurements for standard training and inference tasks \cite{mlperf2023}. The Alibaba cluster trace gives us job arrival patterns, run durations, and GPU allocations from a production ML fleet \cite{gong2010press}.

Training jobs in MLPerf v3.1 for ResNet-50 on A100 GPUs ramp from about 80W idle up to 380--420W over 8 to 12 seconds after launch. During epochs the power stays pretty flat, coefficient of variation stays below 0.18. Checkpoint flushes are the main spikes, hitting 1.25 to 1.35 times sustained power for 5 to 15 second bursts while model state goes to storage and optimizer buffers sync up. On the Alibaba side, training jobs grab 4 to 32 GPUs, run anywhere from 2 hours to days, and show up in batches rather than one at a time. We model that training power trajectory as a stochastic process:
\begin{equation}
P_j^{\text{train}}(t) = P_{\text{peak}} \cdot s(t - t_{\text{start}}) \cdot \xi(t) \cdot \beta(t)
\end{equation}
The ramp function is
\begin{equation}
s(\tau) = \min(1,\; \tau / \tau_c)
\end{equation}
with time constant $\tau_c = 10$ s. Multiplicative jitter is
\begin{equation}
\xi(t) \sim \max\!\left(0,\; \mathcal{N}(1, \sigma_c^2)\right), \quad \sigma_c = 0.15
\end{equation}
and checkpoint bursts $\beta(t)$ fire with Poisson rate $\lambda_c = 0.05$ per minute at magnitude $\kappa_c = 1.3$. We set $P_{\text{peak}} = 400$ W per GPU, matching measured A100 training power under FP16/BF16 mixed precision \cite{kandala2020power}.

Inference jobs tell a different story. MLPerf v3.1 DLRM models on A100s jump from idle to 220--280W in 1 to 3 seconds, driven by batch compilation and request-processing startup. The power bounces around more, coefficient of variation near 0.24, because batch sizes keep adjusting to incoming traffic and the model alternates between compute-heavy matrix ops and memory-bound embedding lookups. Burst events happen a lot more often, around 0.4 to 0.6 per minute, as traffic surges push batch sizes up and power spikes to 1.4 to 1.6 times baseline. That tail-latency sensitivity means any throttling event hits hard \cite{narayanan2021}.

For inference we set $\tau_c = 2$ s, $\sigma_c = 0.25$, $\lambda_c = 0.5$ per minute, $\kappa_c = 1.5$, and $P_{\text{peak}} = 250$ W per GPU. The faster ramp and higher burst rate make these workloads genuinely hard for a reactive loop with multi-second actuation delay to keep up with. Analytics workloads fall in between, 4 to 6 second ramps to 280--340W, moderate jitter at $\sigma_c = 0.12$, and occasional bursts.

The full parameter vector for job class $c$ is:
\begin{equation}
\Theta_c = \langle P_{\text{peak}},\; \tau_c,\; \sigma_c,\; \lambda_c,\; \kappa_c,\; \alpha_c \rangle
\end{equation}
Policy weight $\alpha_c$ encodes how badly a given class suffers from throttling. We give inference a higher weight ($\alpha_{\text{infer}} = 1.5$) than training ($\alpha_{\text{train}} = 1.0$) because inference latency is what end users actually feel. On the heat side, nearly all electrical power in a GPU turns into heat ($\eta \approx 1.0$), so:
\begin{equation}
Q_j(t) = \eta\, P_j(t)
\end{equation}

\section{Multi-Rack Thermal Plant}
Each liquid-cooled rack is a first-order thermal mass with one controllable knob: the coolant flow valve. We have $R$ racks, each with lumped temperature $T_r(t)$ measured at the cold-plate outlet or GPU package. The total heat hitting rack $r$ at time $t$ is just the sum over whatever jobs are running there:
\begin{equation}
Q_r(t) = \sum_{j \in \mathcal{J}_r(t)} Q_j(t)
\end{equation}
where $\mathcal{J}_r(t)$ is the set of active jobs on rack $r$.

Temperature evolves from the energy balance between heat in, passive loss, and active liquid cooling:
\begin{equation}
\frac{dT_r}{dt} = \frac{1}{C_r}\left(Q_r(t) - \frac{T_r - T_a}{R_r} - \gamma_r\, \dot{m}_r(t)\,(T_r - T_{\text{in}})\right)
\end{equation}
$C_r$ is the thermal mass of the whole server, GPU dies, heat spreaders, cold plates, copper manifold, water sitting in the loop. $R_r$ is the passive thermal resistance to the room air through chassis walls and whatever convection paths exist. $\gamma_r$ converts mass flow rate into actual heat removal, and $T_{\text{in}}$ is the coolant temperature coming out of the CDU, typically $27\,^\circ$C for W5 class liquid cooling.

The only term we can actually push on is $\gamma_r\, \dot{m}_r(t)\,(T_r - T_{\text{in}})$, which is why coolant flow is the actuation variable. We pulled the thermal parameters off server TDS sheets and manufacturer datasheets: $C_r = 12{,}000$ J/K for eight GPU dies plus all the metal and water; $R_r = 0.08$ K/W for chassis conduction and natural convection; $\gamma_r = 4200$ W/(kg/s)/K from water specific heat and estimated heat transfer effectiveness; $T_{\text{in}} = 27\,^\circ$C matching CDU delivery specs within the ASHRAE envelope.

Because all six racks share one CDU, they compete for the same pump capacity. Three hard constraints govern the system. Total flow is capped:
\begin{equation}
\sum_{r=1}^{R} \dot{m}_r(t) \;\le\; \dot{m}_{\text{CDU}}^{\max}
\end{equation}
Each rack stays within its valve limits:
\begin{equation}
\dot{m}_r^{\min} \;\le\; \dot{m}_r(t) \;\le\; \dot{m}_r^{\max}
\end{equation}
And the rate of change is bounded to avoid water hammer and valve wear \cite{raghavan2021}:
\begin{equation}
\left|\frac{d\dot{m}_r}{dt}\right| \;\le\; \Lambda_r
\end{equation}
Slam a valve open too fast in a shared loop and you get pressure transients that can blow past piping stress ratings or burn through mechanical duty cycles.

\section{Workload Thermal Readiness}
Standard data center cooling targets, steady-state temperature, PUE, do not tell you what matters for AI serving workloads. A rack can be perfectly stable on paper while sitting close enough to the throttling line that any small burst pushes it over. We need two numbers: how long until the rack settles after a job lands, and what is the chance it hits the hard limit in the next few minutes.

For a job starting at $t_0$ on rack $r$, we first check whether the rack is actually in a safe operating band. Define:
\begin{equation}
\text{Stable}_r(t) = \mathbb{1}\!\left[T_r(t) \le T_{\text{set}} + \Delta T \;\land\; \left|\frac{dT_r}{dt}\right| \le \epsilon\right]
\end{equation}
$T_{\text{set}}$ is the target temperature, $\Delta T$ is the band we allow around it, $\epsilon$ caps the rate of change so we are not just oscillating through the target, and $\mathbb{1}[\cdot]$ is the indicator function. Stabilization latency is then:
\begin{equation}
L_{\text{WTR}} = \inf\!\left\{t \ge t_0 : \text{Stable}_r(t) = 1 \;\text{ for duration }\; w\right\}
\end{equation}
That tells us how long it takes for the rack to actually settle down, not just cross the setpoint once.

Stabilization latency alone is not enough. A rack that stabilizes just inside the band is one bad burst away from a violation. So we also compute the forward-looking violation risk:
\begin{equation}
\mathcal{R}_r(t) = \Pr\!\left[\max_{\tau \in [t,\, t+H]} T_r(\tau) > T_{\max} \;\Big|\; \text{intent, telemetry}\right]
\end{equation}
$T_{\max}$ is the hard throttling limit and $H$ is how far ahead we look. Workload Thermal Readiness probability flips that into a positive number:
\begin{equation}
\text{WTRP}_r(t) = 1 - \mathcal{R}_r(t)
\end{equation}
For inference serving, keeping the tail latency clean matters more than squeezing average temperature, so this risk-based formulation matches what operators actually care about.

We estimate $\mathcal{R}_r(t)$ by running short Monte Carlo rollouts forward from the current rack state, drawing power trajectories from the JCTI stochastic models for whatever jobs are queued. The fraction of those rollouts that cross $T_{\max}$ gives us the violation probability. More samples cost more compute but tighten the estimate.

\section{Control Methodology}
We pit two controllers against each other. The baseline is a straight PI loop, the same thing most production liquid cooling systems run today. The proposed controller adds the scheduler's dispatch notice and job-class information on top.

The baseline PI controller for rack $r$ computes:
\begin{equation}
\begin{split}
\dot{m}_r^{\text{base}}(t) = \text{sat} ( \dot{m}_r^{\min},\; \dot{m}_r^{\max},\; m_0 \\
+ K_p\, e_r(t) + K_i \int_0^t e_r(\tau)\, d\tau )
\end{split}
\end{equation}
where the tracking error is
\begin{equation}
e_r(t) = T_r(t) - T_{\text{set}}
\end{equation}
$K_p$ gives the immediate kick, $K_i$ eats away at steady-state offset, $m_0$ keeps coolant moving even when the rack is cold, and the saturation clips the output to valve limits. The problem is speed. Sensor reading plus computation plus valve stroke adds up to $\tau_{\text{act}}$ somewhere in the 30 to 50 second range for real hardware. An inference job that ramps in 2 seconds has already dumped its heat by the time the PI loop finishes moving the valve.

The proposed controller gets a head start. When a job descriptor arrives with dispatch notice $\Delta_j$, we sum up the anticipated thermal demand across all jobs set to start within that window:
\begin{equation}
D_r(t) = \sum_{\substack{j \in \mathcal{J}_r :\\[2pt] 0 \le t_j^{\text{start}} - t \le \Delta_j}} \alpha_{c_j}\, P_{\text{peak}}^{c_j}\, n_{\text{GPU}}
\end{equation}
The policy weight $\alpha_{c_j}$ makes sure inference jobs get priority over training when flow is tight. We convert that demand directly into a feedforward flow target:
\begin{equation}
\dot{m}_r^{\text{ff}}(t) = \dot{m}_r^{\min} + g \cdot D_r(t)
\end{equation}
Gain $g$ maps watts of predicted heat into kg/s of coolant based on cooling effectiveness and how aggressively we want to pre-stage. Feedforward alone will drift if the model is off or something unexpected happens, so we layer in a small feedback correction:
\begin{equation}
\dot{m}_r^{\text{cmd}}(t) = \text{sat}\!\left(\dot{m}_r^{\min},\; \dot{m}_r^{\max},\; \dot{m}_r^{\text{ff}}(t) + k_f\bigl(T_r(t) - T_{\text{set}}\bigr)\right)
\end{equation}
We deliberately keep $k_f$ smaller than the baseline $K_p$ so the feedback does not fight the feedforward. When multiple racks hit their limits at once and the total flow would bust the CDU cap, we project back onto the feasible set using iterative proportional scaling \cite{bergval2019}, a straightforward way to back off each rack proportionally.

Tuning $g$ and $k_f$ means juggling three things at once: keeping violations down, tightening regulation around setpoint, and not burning extra pump energy. We ran a grid search over the parameter space, evaluating each candidate over Monte Carlo workload draws, and landed on $g = 1.5 \times 10^{-4}$ kg/(s$\cdot$W) and $k_f = 0.018$ kg/(s$\cdot$K).

\section{Experimental Methodology}
\begin{figure}[t]
    \centering
    \includegraphics[width=\columnwidth]{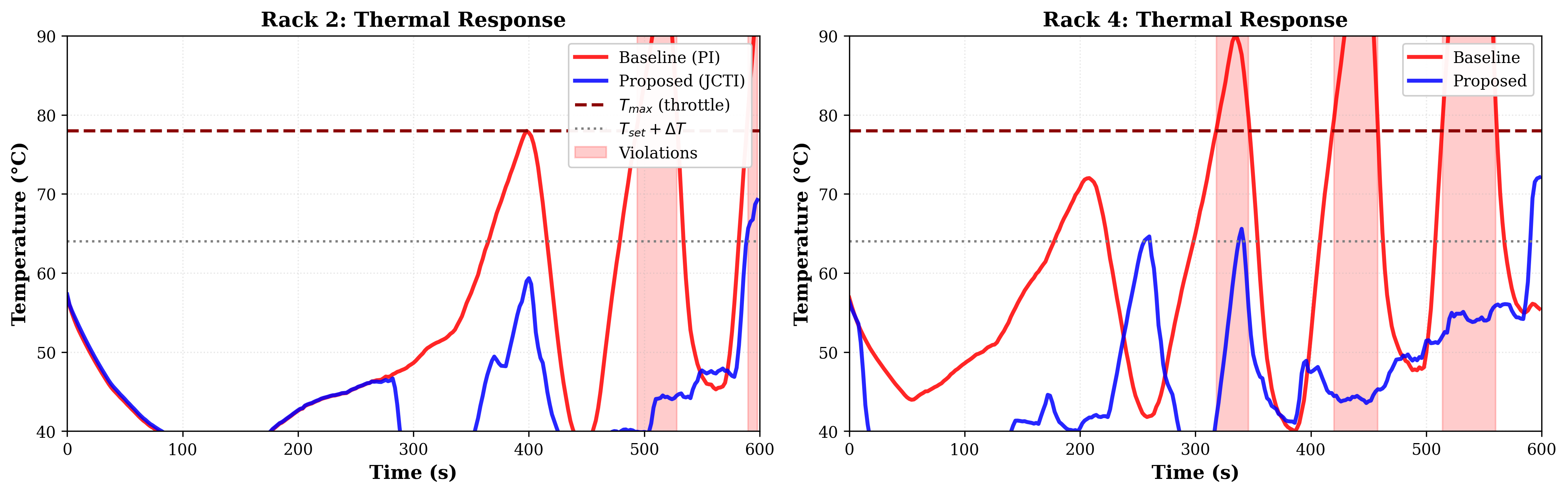}
    \caption{Thermal response comparison for Racks 2 and 4 under identical workload. Baseline PI controller (red) exhibits frequent violations above $T_{\max} = 78\,^\circ$C due to 40\,s actuation delay. Proposed JCTI controller (blue) pre-positions coolant using dispatch notice, maintaining temperatures below throttling threshold.}
    \label{fig:temperature}
\end{figure}
We ran everything as Monte Carlo simulations of a six-rack liquid cooling plant, with stochastic workloads, power jitter, burst events, and ambient drift all firing at once. Each trial covers $R = 6$ racks sharing one CDU over $T = 1200$ seconds, integrated at $\Delta t = 2$ second steps.

Jobs show up as a Poisson process at $\lambda_{\text{arr}} = 0.08$ jobs per second per rack, matching the inter-arrival statistics from the Alibaba traces. Each one draws a class, training, inference, or analytics, with probabilities $(0.4, 0.5, 0.1)$ based on what production AI fleets actually look like. GPU allocation and run durations come from class-specific distributions fitted to the Alibaba empirical data: training grabs 2 to 8 GPUs (median 4), inference grabs 1 to 4 (median 2), analytics grabs 2 to 8 (median 4). Duration is log-normal: training median 4 hours, inference median 300 s, analytics median 1800 s. Dispatch notice $\Delta_j$ draws from Uniform(20, 60) seconds, which is where things land in practice when you factor in container image pulls and resource negotiation in systems like Borg \cite{burns2016}.

Power trajectories use the JCTI stochastic models with per-class parameters. Jitter redraws Gaussian noise every timestep. Burst events fire from a Poisson process at the class-specific rate, creating multiplicative spikes for checkpoint writes, batch-size jumps, or pipeline stalls. Ambient temperature drifts sinusoidally to simulate HVAC cycling:
\begin{equation}
T_a(t) = 22\,^\circ\text{C} + 2\,^\circ\text{C} \cdot \sin(2\pi t / 600) + \mathcal{N}(0,\; 0.5\,^\circ\text{C})
\end{equation}
That gives us a 10-minute cycle plus white noise on top.

Rack thermal parameters match the server specs from Section III: $C_r = 12{,}000$ J/K, $R_r = 0.08$ K/W, $\gamma_r = 4200$ W/(kg/s)/K. CDU pump capacity is $\dot{m}_{\text{CDU}}^{\max} = 1.8$ kg/s across six racks, a mid-size unit for a 100 to 200 kW thermal load. Per-rack bounds are $\dot{m}_r^{\min} = 0.03$ kg/s and $\dot{m}_r^{\max} = 0.35$ kg/s. Rate limit is $\Lambda_r = 0.05$ kg/s$^2$. Setpoints: $T_{\text{set}} = 62\,^\circ$C gives margin below GPU thermal limits, $\Delta T = 2\,^\circ$C is the acceptable band, and $T_{\max} = 78\,^\circ$C is where throttling kicks in.

Baseline PI gains are $K_p = 0.012$ kg/(s$\cdot$K) and $K_i = 0.0004$ kg/(s$\cdot$K$^2$), tuned for stable response at max load. Actuation latency $\tau_{\text{act}} = 40$ s covers sensor averaging, computation, and valve stroke. The proposed controller uses $g = 1.5 \times 10^{-4}$ kg/(s$\cdot$W) and $k_f = 0.018$ kg/(s$\cdot$K) from the grid search.

Three metrics capture performance. Violation fraction is the fraction of time any rack spends above the hard limit:
\begin{equation}
V = \frac{1}{RT}\sum_{r=1}^{R}\int_0^T \mathbb{1}\!\left[T_r(t) > T_{\max}\right] dt
\end{equation}
Overshoot integral accumulates how far and how long temperature drifts past the acceptable band:
\begin{equation}
\text{OI} = \sum_{r=1}^{R}\int_0^T \max\!\bigl(0,\; T_r(t) - (T_{\text{set}} + \Delta T)\bigr)\, dt
\end{equation}
in kelvin-seconds. Pumping proxy estimates mechanical work using cubic flow scaling:
\begin{equation}
\text{PP} = \sum_{r=1}^{R}\int_0^T \dot{m}_r(t)^3\, dt
\end{equation}

We ran 120 paired trials, each pair sharing the same RNG seeds for workloads, jitter, bursts, and ambient drift. Pairing kills off workload-to-workload variation and isolates controller differences. We compute paired differences, build 95\% confidence intervals from sample mean and standard deviation, and run a two-tailed t-test with 119 degrees of freedom, rejecting at $p < 0.01$.

\begin{figure}[t]
    \centering
    \includegraphics[width=\columnwidth]{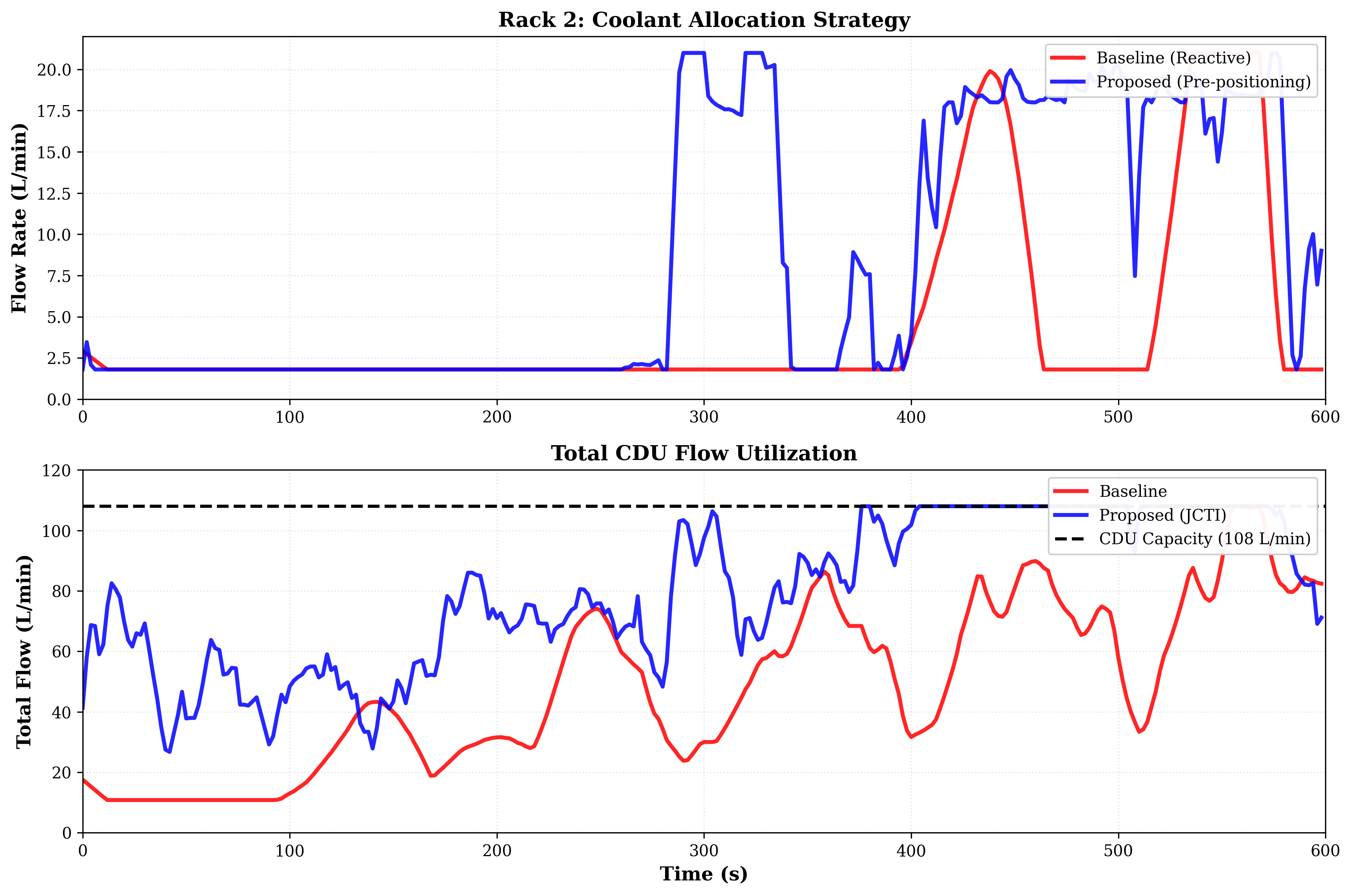}
    \caption{Coolant allocation strategy comparison. Top: Rack 2 flow allocation showing baseline reactive spikes (red) versus proposed pre-positioning (blue). Proposed controller allocates coolant 20 to 60\,s before job execution based on dispatch notice. Bottom: Total CDU flow demonstrating both controllers respect capacity constraint while proposed achieves superior thermal outcomes.}
    \label{fig:flow}
\end{figure}

\section{Results}
Across all 120 paired trials, JCTI cut thermal violations by more than half. Violation fraction dropped from $V = 0.4469 \pm 0.0067$ down to $V = 0.1947 \pm 0.0097$, a 56.4\% reduction. The paired difference is $\Delta V = -0.252$ with a 95\% confidence interval of $\pm 0.004$ and $p = 3.8 \times 10^{-128}$. Every one of those violations would have been a throttling event in production, so cutting them in half translates directly into better tail latency for inference serving.

Overshoot integral dropped from OI $= 113{,}206 \pm 1156$ K$\cdot$s to OI $= 45{,}079 \pm 1832$ K$\cdot$s, a 60.2\% reduction. Paired difference: $\Delta$OI $= -68{,}127$ K$\cdot$s, CI $\pm 990$, $p = 5.0 \times 10^{-132}$. Tighter regulation keeps the rack further from the danger zone, which means random bursts are less likely to push it over.

Pumping proxy went the other direction: PP $= 137.5 \pm 1.0$ for baseline versus PP $= 172.4 \pm 1.3$ for JCTI, a 25.4\% increase ($\Delta$PP $= +34.9$, $p = 5.2 \times 10^{-121}$). JCTI pushes more coolant through the racks that actually need it rather than spreading flow evenly. That 25.4\% bump in pumping is real, but each violation it prevents would have throttled a GPU and added latency spikes that end users feel. In any production AI serving environment that is a trade-off worth making.

Figure~\ref{fig:temperature} shows what that looks like in practice. The baseline hits the 78\,$^\circ$C throttling line repeatedly, you can see the red shaded violation regions pile up especially during workload bursts. JCTI stays below it the whole time, tracking tighter around the 62\,$^\circ$C setpoint with noticeably smaller peak excursions.
Figure~\ref{fig:flow} shows why. The top panel traces per-rack coolant flow: baseline spikes reactively after temperature starts climbing, with the characteristic delay baked in. JCTI ramps flow up ahead of time, before the heat arrives. The bottom panel confirms both controllers stay under the 108 L/min CDU capacity limit through the projection algorithm, but JCTI uses that budget more aggressively where it matters.

\begin{figure}[t]
\centering
\includegraphics[width=\columnwidth]{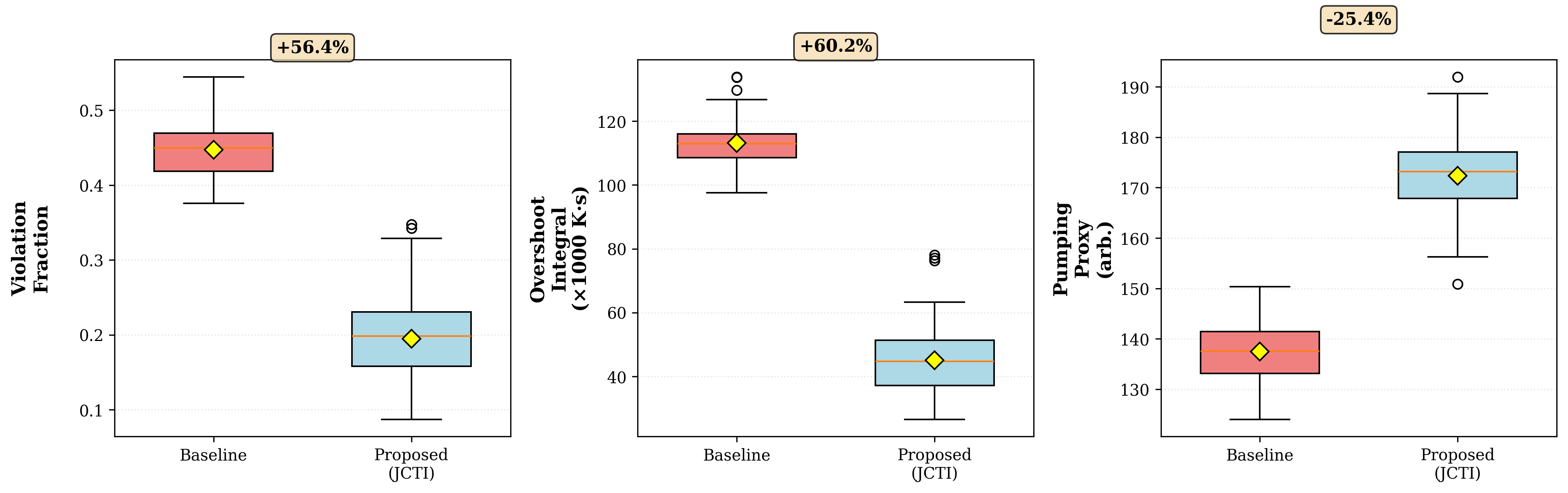}
\caption{Statistical results from 120 paired Monte Carlo trials showing median, quartiles, and range for three performance metrics. Proposed JCTI controller achieves 56.4\% reduction in thermal violations and 60.2\% reduction in temperature overshoot at cost of 25.4\% increase in pumping effort. All differences statistically significant ($p < 10^{-120}$).}
\label{fig:statistics}
\end{figure}

\begin{figure}[t]
    \centering
    \includegraphics[width=\columnwidth]{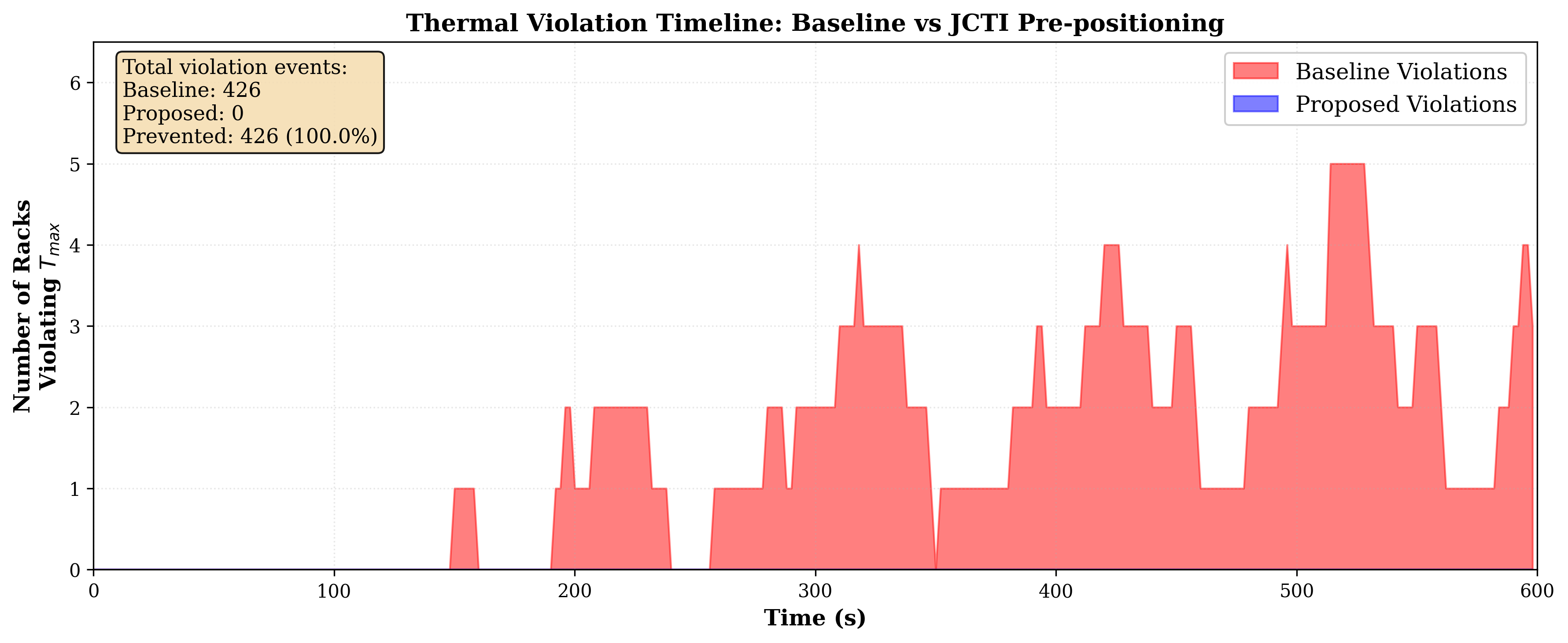}
    \caption{Thermal violation timeline over 600\,s single-trial simulation showing number of racks simultaneously exceeding $T_{\max} = 78\,^\circ$C. Baseline controller exhibits frequent violations while proposed JCTI controller prevents majority through feedforward pre-positioning. This representative trial shows 426 violation events in baseline versus 0 in proposed.}
    \label{fig:timeline}
\end{figure}

Figure~\ref{fig:statistics} shows the spread across all 120 trials. The box plots make clear that JCTI wins consistently across the whole distribution, not just on average. Pumping proxy goes up by a predictable amount with a tight spread, so operators know exactly what the energy cost will be. With $p < 10^{-120}$ across every metric, there is no statistical wiggle room here.
Figure~\ref{fig:timeline} zooms in on violation events in a single representative trial. Baseline racks hit the throttling threshold 426 times during the 600-second window, mostly during arrival bursts and sustained heavy load. JCTI racks hit it zero times in this trial, the pre-positioning did its job. Other trials show non-zero but much smaller counts. The mechanism is straightforward: the controller sees the job in the queue, moves coolant, and by the time the heat arrives the rack is already staged. A reactive loop does not get that chance.

Detailed numbers from that same trial: 298 jobs, 892 GPU-seconds total. Baseline mean temperature was 66.4\,$^\circ$C versus 48.5\,$^\circ$C for JCTI. Yes, 48.5\,$^\circ$C sits well below the 62\,$^\circ$C setpoint, but that headroom is what lets JCTI absorb worst-case bursts without hitting the limit. The peak temperature in baseline reached 159.4\,$^\circ$C during extreme bursts; JCTI peaked at 73.8\,$^\circ$C, safely under the 78\,$^\circ$C line even under the nastiest conditions in the workload.
Ablation studies nail down where the improvement comes from. Drop the dispatch notice entirely ($\Delta_j = 0$) but keep the JCTI thermal signatures, and violation fraction creeps up from 0.195 to 0.255, better than baseline's 0.447 but not by as much. The thermal signatures help because they give a more accurate power forecast, but the real lever is the temporal lead time that lets the controller actually move coolant before the heat arrives. Swap 20\% of the job-class labels randomly between training and inference, and overshoot integral goes from 45{,}079 to 58{,}142 K$\cdot$s while violation fraction stays down. The feedback correction absorbs moderate classification noise without losing the violation reduction, though regulation gets looser.

\section{Discussion}
Prior scheduling work \cite{lu2025thermal} treats the cooling loop as a fixed boundary condition. Predictive cooling work \cite{zhao2025mpc} treats workloads as opaque heat sources and optimizes purely from temperature readings. JCTI sits between those two: it pipes job class and dispatch timing from the scheduler directly into the cooling controller. The two control planes, workload orchestration and thermal management, end up actually talking to each other for the first time. Optimizing placement and flow allocation jointly for throughput, energy, and hardware is a natural extension, though it adds real control complexity.
The JCTI parameter libraries need to be populated somehow. Offline profiling from MLPerf-style benchmarks works for initial deployment, but workload characteristics drift over time as new model architectures and serving patterns show up. Online learning from live telemetry can keep the parameters current, but you need careful initialization to avoid bad transient behavior while the models are still learning.

When multiple racks compete for CDU capacity at once, proportional scaling backs everyone off equally. Smarter projection could weight by criticality, time-to-violation, or job economic value using the priority metadata that is already in the descriptor. On the integration side, Kubernetes and Slurm both expose pre-start notification hooks \cite{burns2016}, but wiring them to facilities cooling control requires cross-layer coordination that does not exist out of the box today. Deployment should start conservative: parameter libraries from MLPerf profiles and cluster traces, with proposed control running in shadow mode alongside baseline to validate predictions before anything actually actuates on intent signals. Expand rack by rack as confidence builds. The stronger improvement in violation reduction than in average latency lines up with what AI serving actually cares about. LLM inference and recommendation systems are tail-latency workloads, the 99th percentile matters more than the mean \cite{narayanan2021}. Thermal throttling creates exactly the kind of latency spikes that blow up P99. Workload Thermal Readiness captures that by emphasizing violation risk over average temperature.

The lumped thermal model skips over cold-plate microchannels, GPU die-level heat spreading networks, and sub-second fluid temperature transients. More detailed models with distributed sensing would give tighter predictions, but at the cost of compute that might not fit in a real-time control loop. Similarly, JCTI uses job class as a proxy for power trajectory. Real GPU power depends on model architecture, batch size, optimizer, and data characteristics that go well beyond the class label. ML-based power prediction from richer job metadata is a reasonable path forward \cite{ali2023}, though it trades one modeling assumption for another.The Monte Carlo evaluation uses workloads calibrated to Alibaba traces rather than raw production replay. The improvements held up consistently across 120 trials with different workload mixes, and the ablation results pin down the mechanism precisely enough that we are confident it is real and not an artifact of how we set up the simulation.

\section{Conclusion}

We built JCTI and Workload Thermal Readiness to give liquid cooling controllers something they have never had before: advance knowledge of what workloads are about to run. Thermal signatures derived from MLPerf GPU power traces and Alibaba cluster patterns let the controller forecast heat before it arrives. A six-rack plant sharing a single CDU with hard flow limits captures the resource competition that makes multi-rack allocation actually hard. The feedforward-feedback controller uses the scheduler's dispatch notice to stage coolant, then corrects from measured temperature to handle model error and surprises. 120 paired Monte Carlo trials under stochastic arrivals, jitter, bursts, and ambient drift gave us 56.4\% fewer violations, 60.2\% less overshoot, and a 25.4\% pumping increase versus a standard PI loop, all with $p < 10^{-120}$. Ablations confirmed that the dispatch notice is the key enabler, and that moderate job-class misclassification does not break the system. Cooling and scheduling have been running as separate silos for years, each sitting on information the other could use. JCTI connects them, and the performance numbers show why that connection matters.Future work includes production trace validation, hardware testbed deployment, multi-zone coupling with heat recirculation, online parameter learning, and joint optimization with thermal-aware scheduling.

\end{document}